\documentclass[12pt,preprint]{aastex}

\usepackage{bm}        
\usepackage{amssymb, amsmath}   
\usepackage{color}

\newcommand{\vect}[1]{\boldsymbol{#1}}

\DeclareMathOperator{\sech}{sech}

\shorttitle{Electron Acceleration during Low-$\beta$ Reconnection}
\shortauthors{Xiaocan Li et al.}
\slugcomment{Accepted for Publication in {\it The Astrophysical Journal Letters}}

\begin{document}
  \title{Nonthermally Dominated Electron Acceleration during Magnetic
  Reconnection in a Low-$\beta$ Plasma}
  \author{Xiaocan Li\altaffilmark{1,2,3}, Fan Guo\altaffilmark{3},
    Hui Li\altaffilmark{3}, Gang Li\altaffilmark{1,2}}
  \altaffiltext{1}{Department of Space Science, University of Alabama in
  Huntsville, Huntsville, AL 35899, USA}
  \altaffiltext{2}{Center for Space Plasma and Aeronomic Research, 
  University of Alabama in Huntsville, Huntsville, AL 35899, USA}
  \altaffiltext{3}{Los Alamos National Laboratory, Los Alamos, NM 87545, USA}

  \begin{abstract}
    By means of fully kinetic simulations, we investigate
    electron acceleration during magnetic reconnection in a nonrelativistic
    proton--electron plasma with conditions similar to solar corona and flares.
    We demonstrate that
    reconnection leads to a nonthermally dominated electron acceleration
    with a power-law energy distribution in the nonrelativistic low-$\beta$
    regime but not in the high-$\beta$ regime, where $\beta$ is the ratio
    of the plasma thermal pressure and the magnetic pressure.
    The accelerated electrons contain most of the dissipated magnetic
    energy in the low-$\beta$ regime.
    A guiding-center current description is used to reveal the role of
    electron drift motions during the bulk nonthermal energization.
    We find that the main acceleration mechanism is a \textit{Fermi}-type
    acceleration accomplished by the particle curvature drift motion
    along the electric field induced by the reconnection outflows.
    Although the acceleration mechanism is similar for different
    plasma $\beta$, low-$\beta$ reconnection drives fast acceleration
    on Alfv\'enic timescales and develops power laws out of thermal distribution.
    The nonthermally dominated acceleration resulting
    from magnetic reconnection in low-$\beta$ plasma may have strong
    implications for the highly efficient electron acceleration in solar
    flares and other astrophysical systems.
  \end{abstract}


  \maketitle

  \section{Introduction}
  Magnetic reconnection is a fundamental plasma process during 
  which the magnetic field restructures itself and converts its energy 
  into plasma kinetic energies (e.g.~\citet{Priest2000Magnetic}).
  It occurs ubiquitously in laboratory,
  space, and astrophysical magnetized plasmas. An important 
  unsolved problem is the acceleration of nonthermal particles 
  in the reconnection region.
  Magnetic reconnection has been suggested as a primary mechanism for
  accelerating nonthermal particles
  in solar flares~\citep{Masuda1994Loop,Krucker2010Measure,Lin2011Energy},
  Earth's magnetosphere~\citep{Oieroset2002Evidence,Fu2011Fermi,Huang2012Electron},
  the sawtooth crash of tokamaks~\citep{Savrukhin2001Generation},
  and high-energy astrophysical 
  systems~\citep{Colgate2001Origin,Zhang2011Internal}.
  In particular, observations of solar flares have revealed
  an efficient particle energization with $10\%-50\%$ of magnetic
  energy converted into energetic electrons and 
  ions~\citep{Lin1976Nonthermal}. The energetic particles usually
  develop a power-law energy distribution that contains energy on the 
  same order of the dissipated magnetic 
  energy~\citep{Krucker2010Measure,Oka2015Electron}.
  Some observations find that the emission has no
  distinguishable thermal component, indicating that most of the electrons
  are accelerated to nonthermal energies~\citep{Krucker2010Measure,
  Krucker2014Particle}. This efficient production of energetic
  particles poses a challenge to current theories of particle
  acceleration.

  Particle acceleration associated with reconnection has been studied in
  reconnection-driven turbulence~\citep{Miller1997Stochastic}, 
  at shocks in the outflow region~\citep{Tsuneta1998Fermi,Guo2012Particle},
  and in the reconnection
  layer~\citep{Drake2006Electron, Fu2006Process, Oka2010Electron,
    Kowal2012Particle, Guo2014Formation, Zank2014Particle}.
  Previous kinetic simulations have examined various acceleration
  mechanisms during reconnection, including the \textit{Fermi}-type mechanism
  in magnetic islands (flux ropes in three-dimensional
  simulations;~\citet{Drake2006Electron, Guo2014Formation})
  and direct acceleration in the diffusion
  region~\citep{Pritchett2006Relativistic, Huang2010Mechanisms}.
  Most simulations focus on regimes with plasma $\beta\geq0.1$
  with no obvious power-law distributions observed.
  It was argued that particle loss from the 
  simulation domain is important for developing a power-law
  distribution~\citep{Drake2010Magnetic}. 
  Early simulations of relativistic reconnection showed power-law distributions
  within the X-region~\citep{Zenitani2001Generation}.
  Recent kinetic simulations with a highly magnetized
  ($\sigma=\frac{B^2}{4\pi n_em_ec^2}\gg1$)
  pair plasma found global power-law distributions without the particle loss,
  although the loss mechanism may be important
  in determining the spectral index~\citep{Guo2014Formation,Guo2015Particle}.
  It is unknown whether or not this is valid for reconnection in a
  nonrelativistic proton--electron plasma, since its property is
  different from the relativistic reconnection~\citep{Liu2015Scaling}.

  Motivated by the results of relativistic reconnection,
  here, we report fully kinetic simulations of magnetic
  reconnection in a nonrelativistic proton--electron plasma with a range
  of electron and ion $\beta_e=\beta_i=0.007-0.2$. The low-$\beta$ regime
  was previously relatively unexplored due to various numerical challenges.
  We find that reconnection in
  the low-$\beta$ regime drives efficient energy conversion and
  accelerates electrons into a power-law distribution $f(E)\sim E^{-1}$.
  At the end of the low-$\beta$ cases,
  more than half of the electrons in number and $90\%$ in energy are in 
  the nonthermal electron population.
  This strong energy conversion and particle acceleration led to
  a post-reconnection region with the kinetic energy of energetic particles
  comparable to magnetic energy.
  Since most electrons are magnetized in the low-$\beta$
  plasma, we use a guiding-center drift description to demonstrate that the main
  acceleration process is a \textit{Fermi}-type mechanism through the particle
  curvature drift motion along the electric field induced by fast plasma flows.
  The development of power-law distributions is consistent with the
  analytical model~\citep{Guo2014Formation}.
  The nonthermally dominated energization may help explain the
  efficient electron acceleration in the low-$\beta$ plasma environments,
  such as solar flares and other astrophysical reconnection sites.

  In Section~\ref{sec:numerical}, we
  describe the numerical simulations. In Section~\ref{sec:results}, we
  present simulation results and discuss the conditions for
  the development of power-law distributions.
  We discuss and conclude the results in Section~\ref{sec:discussion}.
  
  \section{Numerical simulations}
  \label{sec:numerical}
  The kinetic simulations are carried out using the VPIC
  code~\citep{Bowers2008PoP}, which solves Maxwell's equations and
  follows particles in a fully relativistic manner. The
  initial condition is a force-free current sheet with a magnetic field
  $\vect{B} = B_0\tanh(z/\lambda)\hat{x}+B_0\sech(z/\lambda)\hat{y}$,
  where $\lambda=d_i$ is the half thickness of the layer.
  Here, $d_i$ is the ion inertial length.
  The plasma consists of protons and electrons with a mass ratio
  $m_i/m_e$ = 25.
  The initial distributions for both electrons
  and protons are Maxwellian with uniform density $n_0$ and temperature
  $kT_i=kT_e=0.01m_ec^2$. 
  A drift velocity for electrons $U_e$ is added to represent
  the current density that satisfies the Ampere's law.
  The initial electron and ion $\beta_e=\beta_i=8\pi n_0kT_e/B_0^2$
  are varied by changing $\omega_{pe}/\Omega_{ce}$,
  where $\omega_{pe}=\sqrt{4\pi n_0e^2/m_e}$ is the 
  electron plasma frequency
  and $\Omega_{ce}=eB_0/(m_ec)$ is the electron gyrofrequency.
  Quantities $\beta_e=0.007$, 0.02, 0.06, and 0.2
  correspond to $\omega_{pe}/\Omega_{ce} = 
  0.6$, 1, $\sqrt{3}$ and $\sqrt{10}$, respectively.
  The domain sizes are $L_x\times L_z=200d_i\times100d_i$.
  We use $N_x\times N_z=4096\times2048$ cells with 200 particles
  per species per cell.
  The boundary conditions are periodic
  along the $x$-direction, perfectly conducting
  boundaries for fields and reflecting boundaries for particles along
  the $z$-direction. A long wavelength perturbation is added to induce
  reconnection~\citep{Birn2001Geospace}.

  \section{Simulation results}
  \label{sec:results}

\begin{figure}[htbp]
    \centering
    \includegraphics[width=3.4in]{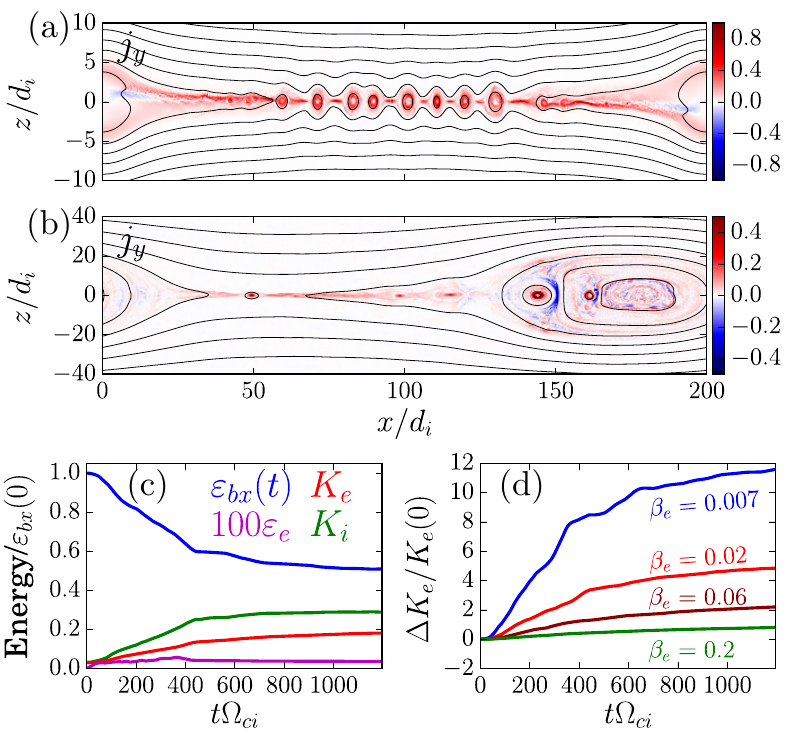}
    \caption{\label{fig:fig1} Out-of-plane current density for the
    case with $\beta_e=0.02$ at
    (a) $t\Omega_{ci}=62.5$ and (b) $t\Omega_{ci}=400$. (c) The energy evolution
    for the $\beta_e=0.02$ case.
    $\varepsilon_{bx}(t)$ is the magnetic energy of the reconnecting component.
    $\varepsilon_e$ is the electric energy. $K_i$ and $K_e$ are ion
    and electron kinetic energies normalized by $\varepsilon_{bx}(0)$, respectively.
    (d) The ratio of electron energy gain $\Delta K_e$ to the initial $K_e$
    for different $\beta_e$.
  }
  \end{figure}

  Under the influence of the initial perturbation, the current sheet quickly
  thins down to a thickness of $\sim d_e$ (electron inertial length
  $c/\omega_{pe}$) that is unstable to the secondary
  tearing instability~\citep{Daughton2009Transition, Liu2013Bifurcated}.
  Fig.~\ref{fig:fig1}(a) and (b) show the evolution of the out-of-plane
  current density. The reconnection layer breaks and generates a chain of
  magnetic islands that interact and coalesce with each other. The largest
  island eventually grows comparable to the system size and the reconnection
  saturates at $t\Omega_{ci}\sim800$.
  Fig.~\ref{fig:fig1}(c) shows the time evolution of the magnetic energy
  in the $x$-direction (the reconnecting component) $\varepsilon_{bx}$
  and the kinetic
  energy of electrons $K_e$ and ions $K_i$ for the case with $\beta_e=0.02$,
  respectively. Throughout the simulation, 40\% of the initial $\varepsilon_{bx}$
  is converted into plasma kinetic energy. Of the converted energy, 38\%
  goes into electrons and 62\% goes into ions.
  We have carried out simulations with larger domains (not shown)
  to confirm that the energy conversion is still efficient and weakly depends
  on system size.
  Since the free magnetic energy overwhelms the initial kinetic energy,
  particles in the reconnection region are strongly energized. Eventually,
  $K_e$ and $K_i$ are 5.8 and 9.4 times their
  initial values, respectively. Fig.~\ref{fig:fig1}(d) shows
  the ratio of the electron energy gain $\Delta K_e$
  to the initial electron energy $K_e(0)$ for different cases.
  While the $\beta_e=0.2$ case
  shows only mild energization, cases with lower $\beta_e$ give stronger
  energization as the free energy increases.

\begin{figure}[htbp]
    \centering
    \includegraphics[width=3.4in]{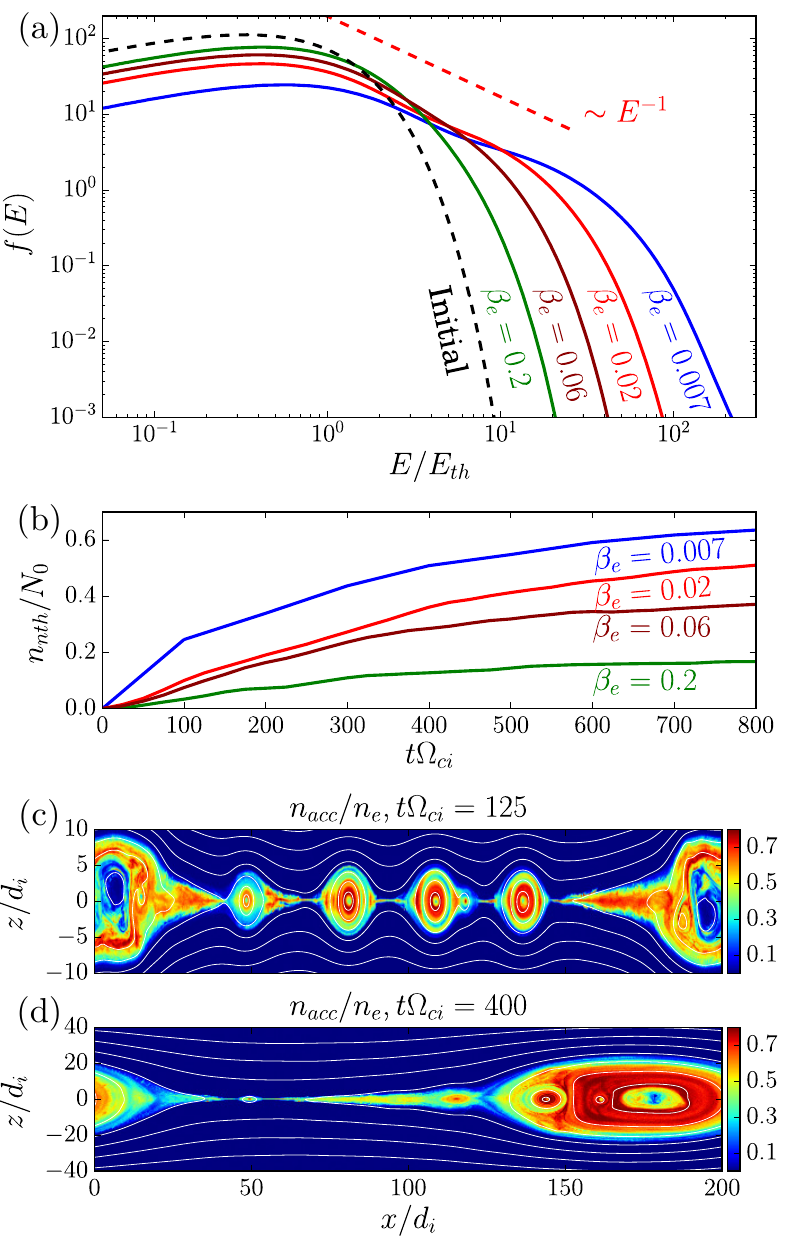}
    \caption{\label{fig:fig2}
      (a) Electron energy spectra $f(E)$ at $t\Omega_{ci}=800$ for
      different $\beta_e$. The electron energy $E$ is normalized
      to the initial thermal energy $E_{th}$.
      The black dashed line is the initial thermal distribution.
      (b) Time evolution of the fraction of nonthermal
      electrons for different initial $\beta_e$.
      $n_{nth}$ is the number of nonthermal
      electrons obtained by subtracting the fitted
      thermal population from the whole particle distribution.
      The fraction of electrons with energies larger than three times
      the initial thermal energy at (c) $t\Omega_{ci}=125$ and
      (d) $t\Omega_{ci}=400$.
  }
  \end{figure}

  The energy conversion drives strong nonthermal electron acceleration.
  Fig.~\ref{fig:fig2}(a) shows the final electron energy spectra over
  the whole simulation domain for the four cases.
  More electrons are accelerated to high energies for lower-$\beta$ cases,
  similar to earlier simulations~\citep{Bessho2010Fast}.
  More interestingly, in the cases with $\beta_e=0.02$ and $0.007$,
  the energy spectra develop a power-law-like tail $f(E)\sim E^{-p}$
  with the spectral index $p\sim1$.
  This is similar to results from relativistic
  reconnection~\citep{Guo2014Formation,Guo2015Particle}.
  We have carried out one simulation with $m_i/m_e=100$ and $\beta_e=0.02$
  and find a similar electron spectrum.
  In contrast, the case with $\beta_e=0.2$ does not show any obvious
  power-law tail, consistent with earlier
  simulations~\citep{Drake2010Magnetic}.
  The nonthermal population dominates the distribution
  in the low-$\beta$ cases. For example, when we subtract
  the thermal population by fitting the low-energy distribution
  as Maxwellian, the nonthermal tail in the $\beta_e=0.02$ case
  contains $55\%$ of the electrons and $92\%$ of the total electron energy.
  The power-law tail breaks at energy $E_b\sim 10E_{th}$ for $\beta_e=0.02$,
  and extends to a higher energy for $\beta_e=0.007$.
  Fig.~\ref{fig:fig2}(b) shows the fraction of 
  nonthermal electrons for different cases. 
  For $\beta_e=0.007$, the nonthermal fraction
  goes up to $66\%$, but it decreases to $17\%$ for $\beta_e=0.2$.
  Fig.~\ref{fig:fig2}(c) and (d)
  show $n_{acc}/n_e$ at $t\Omega_{ci}=125$ and 400 for
  the case with $\beta_e=0.02$,
  where $n_{acc}$ is the number density
  of accelerated electrons with energies larger than three times their
  initial thermal energy, and $n_e$ is the total electron number density.
  The fraction of energetic electrons is over 40\% and up to 80\%
  inside the magnetic islands and reconnection exhausts, indicating a bulk
  energization for most of electrons in the reconnection layer.
  The energetic electrons will eventually
  be trapped inside the largest magnetic island.
  The nonthermally dominated distribution contains most of
  the converted magnetic energy, indicating that energy conversion 
  and particle acceleration are intimately related.

\begin{figure}[htbp]
    \centering
    \includegraphics[width=3.4in]{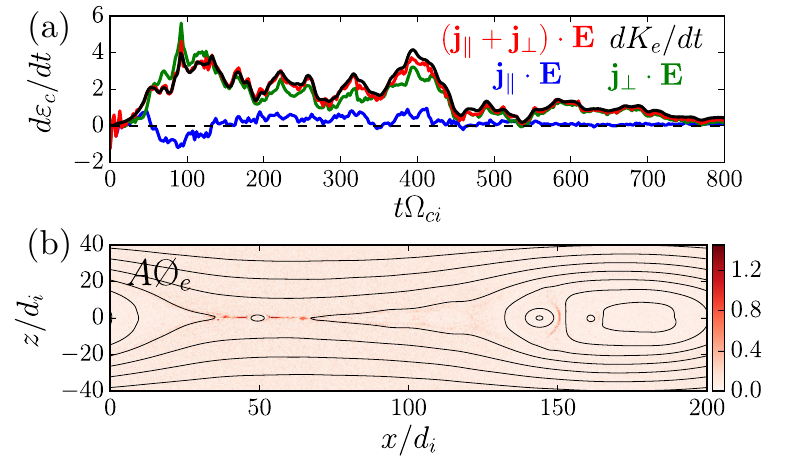}
    \caption{\label{fig:fig3}
      (a) Energy conversion rate $d\varepsilon_c/dt$ for electrons
      through the parallel and
      perpendicular directions with respect to the local magnetic field,
      compared with the energy change rate of electrons $dK_e/dt$ for
      the case with $\beta_e=0.02$.
      The shown values are integrals over the whole simulation domain.
      (b) Electron pressure agyrotropy $A\text{\O}_e$ at $t\Omega_{ci}=400$
      in the same case. See the text for details.
  }
  \end{figure}

  To study the energy conversion, Fig.~\ref{fig:fig3}(a)
  shows the energy conversion rate $d\varepsilon_c/dt$ from the magnetic
  field to electrons through directions parallel and perpendicular
  to the local magnetic field.
  We define $d\varepsilon_c/dt=\int_\mathcal{D}\vect{j}'\cdot\vect{E}dV$,
  where $\mathcal{D}$ indicates the simulation domain and $\vect{j}'$ is
  $\vect{j}_\parallel$ or $\vect{j}_\perp$.
  We find that energy conversion from the perpendicular directions
  gives $\sim90\%$ of the electron energy gain.
  By tracking the trajectories (not shown) of a large number of
  accelerated electrons, we find various acceleration processes in the
  diffusion region, magnetic pile-up region, contracting islands, and
  island coalescence regions~\citep{Hoshino2001Suprathermal,
    Hoshino2005Electron, Drake2006Electron, Fu2006Process,
  Huang2010Mechanisms, Oka2010Electron, Dahlin2014Mechanisms,
  Guo2014Formation}. The dominant
  acceleration is by particles bouncing back and forth through a \textit{Fermi}-like
  process accomplished by particle drift motions
  within magnetic structures (Li et al. 2015, in preparation).
  To reveal the role of particle drift motions,
  we use a guiding-center drift description to
  study the electron energization for the $\beta_e=0.02$ case.
  The initial low $\beta$ guarantees that this is a good approximation
  since the typical electron gyroradius $\rho_e$ is smaller than the
  spatial scale of the field variation ($\sim d_i$).

  By ensemble averaging the particle gyromotion and drift motions,
  the perpendicular current density for a single species can be expressed
  as~\citep{Parker1957Newtonian, Blandford2014Cosmic}
  \begin{equation}
    \vect{j}_\perp=P_\parallel
    \frac{\vect{B}\times(\vect{B}\cdot\nabla)\vect{B}}{B^4}+
    P_\perp\left(\frac{\vect{B}}{B^3}\right)\times\nabla B\nonumber
    -\left[\nabla\times\frac{P_\perp\vect{B}}{B^2}\right]_\perp+
    \rho\frac{\vect{E}\times\vect{B}}{B^2}+
    \rho_m\frac{\vect{B}}{B^2}\times\frac{d\vect{u}_E}{dt}
    \label{equ:jperp}
  \end{equation}
  using a gyrotropic pressure tensor
  $\mathcal{P}=P_\perp\mathcal{I}+(P_\parallel-P_\perp)\vect{b}\vect{b}$, where
  $P_\parallel\equiv m_e\int fv_\parallel^2d\vect{v}$ and
  $P_\perp\equiv0.5m_e\int fv_\perp^2d\vect{v}$, $\rho$ is the particle
  charge density, and $\rho_m$ is the particle mass density.
  The terms on the right are due to curvature drift,
  $\nabla B$ drift, magnetization, $\vect{E}\times\vect{B}$ drift, and
  polarization drift, respectively. The expression is simplified as
  $\vect{j}_\perp=\vect{j}_c+\vect{j}_g+\vect{j}_m+
  \vect{j}_{\vect{E}\times\vect{B}}+\vect{j_p}$,
  in which $\vect{j}_{\vect{E}\times\vect{B}}$ has no direct contribution
  to the energy conversion. This gives an accurate description for 
  $\vect{j}_\perp$ if the pressure tensor is gyrotropic.
  To confirm this, we calculate the
  electron pressure agyrotropy $A\text{\O}_e\equiv
  2\frac{|P_{\perp e1}-P_{\perp e2}|}{P_{\perp e1}+P_{\perp e2}}$,
  where $P_{\perp e1}$ and $P_{\perp e2}$ are the two pressure eigenvalues
  associated with eigenvectors perpendicular to the mean magnetic field
  direction~\citep{Scudder2008Illuminating}. $A\text{\O}_e$ measures
  the departure of the pressure tensor from cylindrical symmetry about the
  local magnetic field. It is zero when the local particle
  distribution is gyrotropic.
  Fig.~\ref{fig:fig3}(b) shows that the regions with nonzero $A\text{\O}_e$
  are localized to X-points.
  The small $A\text{\O}_e$ indicates that the electron distributions
  are nearly gyrotropic in most regions.
  Therefore, the drift description is a good approximation for electrons
  in our simulations even without an external guide field, which is required
  for this description in a high-$\beta$ plasma~\citep{Dahlin2014Mechanisms}.
  
\begin{figure}[htbp]
    \centering
    \includegraphics[width=3.4in]{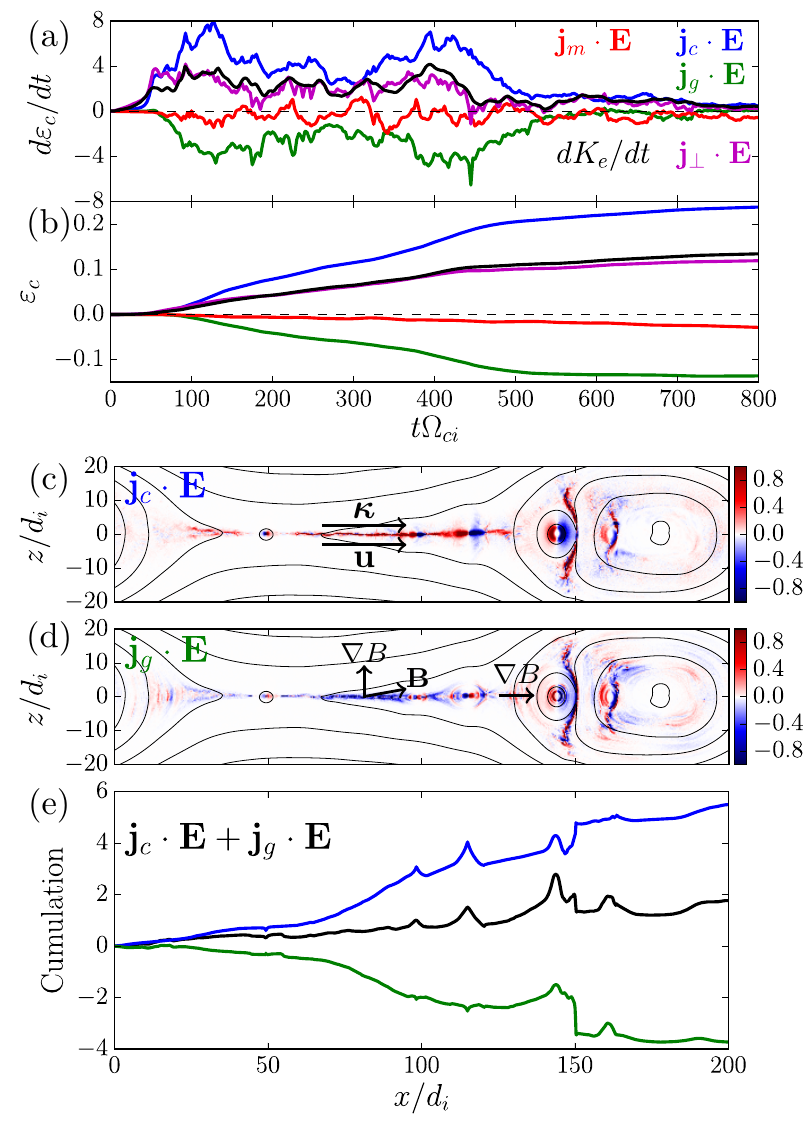}
    \caption{\label{fig:fig4}
      Analysis using a drift description for the case
      with $\beta_e=0.02$. (a) The energy conversion rate due to
      different types of current terms, compared with the electron
      energy change rate $dK_e/dt$.
      $\vect{j}_c\cdot\vect{E}$, $\vect{j}_g\cdot\vect{E}$, and 
      $\vect{j}_m\cdot\vect{E}$ represent energy conversion due to curvature 
      drift, $\nabla B$ drift, and magnetization, respectively.
      (b) The converted magnetic energy due to various terms in (a),
      normalized to the initial magnetic energy of
      the reconnecting component $\varepsilon_{bx}(0)$.
      (c) Color-coded contours of energy conversion rate due to curvature
      drift at $t=400\Omega_{ci}^{-1}$.
      $\bm{\kappa}$ and $\vect{u}$ indicate the directions of the magnetic
      field curvature and the bulk flow velocity.
      (d) Color-coded contours of energy conversion rate due 
      to $\nabla B$ drift at $t=400\Omega_{ci}^{-1}$.
      $\vect{B}$ and $\nabla B$ indicate the directions of the magnetic field
      and the gradient of $|\vect{B}|$.
      Both $\vect{j}_c\cdot\vect{E}$ and $\vect{j}_g\cdot\vect{E}$
      are normalized to the $0.002n_0m_e c^2\omega_{pe}$.
      (e) The cumulation of $\vect{j}_c\cdot\vect{E}$ (blue) and
      $\vect{j}_g\cdot\vect{E}$ (green) along the $x$-direction. The black line
      is the sum of these two.
    }
  \end{figure}

  Fig.~\ref{fig:fig4}(a) and (b) show time-dependent $d\varepsilon_c/dt$
  and $\varepsilon_c$ from different current terms,
  where $\varepsilon_c=\int_0^t(d\varepsilon_c/dt)dt$.
  The contribution from polarization current and parallel current 
  are small and not shown.
  The curvature drift term is a globally dominant term
  of $\vect{j}_\perp\cdot\vect{E}$,
  the $\nabla B$ term gives a net cooling, and
  the magnetization term is small compared to these two.
  Fig.~\ref{fig:fig4}(c) shows the spatial distribution of
  $\vect{j}_c\cdot\vect{E}$.
  When the flow velocity $\vect{u}$ is along
  the magnetic field curvature $\bm{\kappa}$ due to tension force,
  $\vect{j}_c\cdot\vect{E}\approx
  (P_\parallel\vect{B}\times\boldsymbol{\kappa}/B^2)\cdot(-\vect{u}\times\vect{B})>0$.
  These regions are a few $d_i$ along the $z$-direction, but over $50 d_i$
  along the $x$-direction. The overall effect of $\vect{j}_c\cdot\vect{E}$ is
  a strong electron energization.
  Fig.~\ref{fig:fig4}(d) shows that $\vect{j}_g\cdot\vect{E}$ is negative in
  most regions because the strong $\nabla B$ is along the direction
  out of the reconnection exhausts. Then,
  $\vect{j}_g\cdot\vect{E}\sim(\vect{B}\times
  \nabla B)\cdot(-\vect{u}\times\vect{B})<0$.
  Note at some regions, $\vect{j}_g\cdot\vect{E}$ can give strong acceleration.
  Fig.~\ref{fig:fig4}(e) shows the cumulation of the $\vect{j}_c\cdot\vect{E}$
  and $\vect{j}_g\cdot\vect{E}$ along the $x$-direction.
  In the reconnection exhaust region($x=60-115d_i$), $\vect{j}_c\cdot\vect{E}$
  is stronger
  than $\vect{j}_g\cdot\vect{E}$, so the electrons can be efficiently
  accelerated when going through these regions. In the pile-up
  region($x=120-140d_i$), $\bm{\kappa}$, $\nabla B$ and $\vect{u}$
  are along the same direction, so both
  terms give electron energization. In the island coalescence
  region($x\sim 150d_i$), $\vect{j}_c\cdot\vect{E}$ gives electron heating, while
  $\vect{j}_g\cdot\vect{E}$ gives strong electron cooling. Although the net
  effect is electron cooling, island coalescence can be efficient in
  accelerating electrons to the highest energies~\citep{Oka2010Electron}.

  It has been shown that the curvature drift acceleration 
  in the reconnection region corresponds to
  a \textit{Fermi}-type mechanism~\citep{Dahlin2014Mechanisms, Guo2014Formation,
  Guo2015Particle}. To develop a power-law energy distribution for the
  \textit{Fermi} acceleration mechanism, the characteristic acceleration time
  $\tau_{acc}=1/\alpha$ needs to be smaller than the particle injection time
  $\tau_{inj}$~\citep{Guo2014Formation, Guo2015Particle}, 
  where $\alpha=(1/\varepsilon)(\partial\varepsilon/\partial t)$, and 
  $\partial\varepsilon/\partial t$ is the energy change rate of particles.
  To estimate the ordering of acceleration rate
  from the single-particle drift motion, consider the
  curvature drift velocity 
  $\vect{v}_c=v^2_\parallel\vect{B}\times\boldsymbol{\kappa}/(\Omega_{ce}B)$
  in a curved field where $R_c=|\vect{\boldsymbol{\kappa}}|^{-1}$,
  so the time for a particle to cross this region is 
  $\sim R_c/v_\parallel$ and the electric field is mostly induced by
  the Alfv\'enic plasma flow  $\vect{E}\sim-\vect{v}_A\times\vect{B}/c$.
  The energy
  gain in one cycle is $\delta\varepsilon\sim mv_Av_\parallel$.
  The time for a particle to cross the island is $L_{island}/v_\parallel$.
  Then, the acceleration rate
  $\partial\varepsilon/\partial t\sim\varepsilon v_A/L_{island}$ for a
  nearly isotropic distribution.
  The characteristic acceleration time $\tau_{acc}\sim L_{island}/v_A$.
  Taking $L_{island}\sim50d_i$ and $v_A\sim 0.2c$,
  the acceleration time $\tau_{acc}\sim250\Omega_{ci}^{-1}$.
  The actual acceleration time may be longer because the
  outflow speed will decrease from $v_A$ away from the X-points,
  and the $\nabla B$ term gives a non-negligible cooling effect.
  Our analysis has also found that pre-acceleration and trapping
  effects at the X-line region can lead to more efficient electron
  acceleration by the \textit{Fermi} mechanism and are worthwhile to
  investigate further~\citep{Hoshino2005Electron, Egedal2015Double, Huang2015Electron}.
  Taking the main energy release phase as the injection time
  $\tau_{inj}\sim800\Omega_{ci}^{-1}$, the estimated value of
  $\tau_{inj}/\tau_{acc}\sim 3.2$, well above the threshold.
  For the case with $\beta_e=0.2$, the ratio $\tau_{inj}/\tau_{acc}\sim0.32<1$,
  so there is no power-law energy distribution.

  \section{Discussion and conclusion}
  \label{sec:discussion}

  Nonthermal power-law distributions have rarely been
  found in previous kinetic simulations of nonrelativistic
  magnetic reconnection~\citep{Drake2010Magnetic}.
  We find that two essential conditions are required for producing power-law
  electron distribution.
  The first is that the domain should be large
  enough to sustain reconnection for a sufficient duration.
  A power-law tail develops as the acceleration
  accumulates long enough ($\tau_{inj}/\tau_{acc}>1$).
  The second condition is that plasma $\beta$ must be low to form a
  nonthermally dominated power-law distribution by providing
  enough free energy ($\propto1/\beta$) for nonthermal electrons.
  Assuming 10\% of magnetic energy is
  converted into nonthermal electrons with spectral index $p=1$,
  one can estimate that $\beta_e$ is about 0.02
  for half of the electrons to be accelerated
  into a power law that extends to $10E_{th}$. This agrees well
  with our simulation.  
  We point out that a loss mechanism or radiation cooling
  can affect the final power-law index~\citep{Fermi1949Origin, Guo2014Formation}
  of nonthermal electrons. Consequently, including
  loss mechanisms in a large three-dimensional open system is
  important, for example, to explain the observed power-law index
  in solar flares and other astrophysical processes.
  Another factor that may influence our results is the presence 
  of an external guide field $B_g$. Our preliminary analysis has
  shown that the \textit{Fermi} acceleration dominates when $B_g\lesssim B_0$.
  The full discussion for the cases including the
  guide field will be reported in another publication 
  (Li et al. 2015, in preparation). A potentially important issue is the
  three-dimensional instability, such as kink instability 
  that may strongly influence the results. Unfortunately, the
  corresponding three-dimensional simulation is beyond the
  available computing resources. We note that results from 
  three-dimensional simulations with pair plasmas have shown
  development of strong kink instability but appear to have no 
  strong influence on particle
  acceleration~\citep{Guo2014Formation, Sironi2014Relativistic}.
  The growth rate of the kink instability can be much less
  than the tearing instability for a high mass ratio
  \citep{Daughton1999Unstable}, and therefore
  the kink instability may be even less important for electron 
  acceleration in a proton--electron plasma.

  In our simulations, the low-$\beta$ condition is achieved by increasing
  magnetic field strength (or equivalently decreasing density). We have
  carried out low-$\beta$ simulations with the same magnetic field but
  lower temperature and found a similar power-law
  distribution (Li et al. 2015, in preparation).
  
  The energy partition between electrons and protons shows that
  more magnetic energy is converted into protons. For simulations 
  with a higher mass ratio $m_i/m_e=100$, the energetic electrons still
  develop a power-law distribution, and the fraction of electron 
  energy to the total plasma energy is about 33\%,
  indicating that the energy conversion and electron acceleration are
  still efficient for higher mass ratios.
  Our results show that ions also develop a power-law energy
  spectrum for low-$\beta$ cases and the curvature drift acceleration is
  the leading mechanism. However, the ion acceleration has 
  a strong dependence on the mass ratio $m_i/m_e$ for our relatively
  small simulation domain ($\sim100d_i$).
  We therefore defer the study of ion acceleration
  to a future work (Li et al., 2015, in preparation).

  The energetic electrons can generate observable X-ray emissions.
  As nonthermal electrons are mostly concentrated inside the magnetic
  islands, the generated hard X-ray flux can be strong enough to
  be observed during solar flares in the above-the-loop-top
  region~\citep{Masuda1994Loop, Krucker2010Measure} and the reconnection
  outflow region~\citep{Liu2013Plasmoid}. The nonthermal electrons may also
  account for the X-ray flares in the accretion disk
  corona~\citep{Galeev1979Structured, Haardt1994Model, Li1997Electron}.
  
  In summary, we find that in a nonrelativistic low-$\beta$
  proton--electron plasma, magnetic reconnection
  is highly efficient at converting the free energy stored in a
  magnetic shear into plasma kinetic energy and accelerating electrons
  into nonthermal energies. The nonthermal electrons contain more than
  half of the total electrons, and their distribution resembles
  power-law energy spectra with spectral index $p\sim1$ when particle loss
  is absent. This is in contrast to the high-$\beta$ case, where no
  obvious power-law spectrum is observed. It is important to emphasize that
  the particle acceleration discussed here is distinct from
  the acceleration by shocks, where the nonthermal population contains
  only about 1\% of particles~\citep{Parker2012Particle}.
  
  \acknowledgments
  We gratefully thank William Daughton for providing access
  to the VPIC code and for useful discussions. We also acknowledge 
  the valuable discussions with Andrey Beresnyak and Yi-Hsin Liu. 
  This work was
  supported by NASA Headquarters under the NASA Earth and Space Science
  Fellowship Program-Grant NNX13AM30H and by the DOE through the
  LDRD program at LANL and DOE/OFES support to LANL in 
  collaboration with CMSO. Simulations were performed with LANL
  institutional computing.

\end{document}